\documentclass[
aps,pra,
reprint,
superscriptaddress,
amsmath,
amssymb
]{revtex4-1}

\usepackage[utf8]{inputenc}
\usepackage{times}
\usepackage{microtype}
\usepackage{graphicx}
\usepackage{upgreek}
\usepackage{bm}
\usepackage{xcolor}
\usepackage[normalem]{ulem}
\usepackage{placeins}
\usepackage{dcolumn}

\usepackage{filecontents}
\usepackage[T1]{fontenc}
\usepackage{csquotes}
\usepackage[english]{babel}
\usepackage{centernot}
\usepackage{mathtools}
\usepackage{ifthen}
\usepackage{tikz}
\usetikzlibrary{calc}
\usetikzlibrary{decorations.pathreplacing}
\usetikzlibrary{decorations}
\usetikzlibrary{positioning}
\usepackage{pgfplots}
\usepackage{xxcolor}
\definecolor{structure}{rgb}{0.23,0.4,0.7}
\usepackage[normalem]{ulem}

\newtheorem{definition}{Definition}

\newtheorem{axiom}{Axiom}

\newsavebox{\blocksavebox}

\definecolor{niceblue}{rgb}{0.33,0.5,0.8}

\newcommand{\refsub}[2]{\hyperref[#1]{\ref*{#1}#2}}

\newcommand{\bra}[1]{\langle #1|}
\newcommand{\ket}[1]{|#1\rangle}

\newcommand{\norm}[2][]{
  \ifthenelse{\equal{#1}{}}
    {\left\| {#2} \right\|}
    {\ifthenelse{\equal{#1}{uinv}}
      {\left\vert\kern-0.25ex\left\vert\kern-0.25ex\left\vert {#2} \right\vert\kern-0.25ex\right\vert\kern-0.25ex\right\vert}
      {\left\| {#2} \right\|_{#1}}
    }
}

\newcommand{\taverage}[2][]{
  \ifthenelse{\equal{#1}{}}
  {\overline{#2}}
  {\overline{#2}^{#1}}
}

\newcommand{\tracedistance}[3][]{
  \ifthenelse{\equal{#2}{}}
  {\ifthenelse{\equal{#3}{}}
    {\mathcal{D}_{#1}}{}
  }{
    \ifthenelse{\equal{#1}{}}
    {\mathchoice{\operatorname{\mathcal{D}}\left(#2,#3\right)}{\operatorname{\mathcal{D}}(#2,#3)}{\operatorname{\mathcal{D}}(#2,#3)}{\operatorname{\mathcal{D}}(#2,#3)}}
    {\mathchoice{\operatorname{\mathcal{D}}_{#1}\left(#2,#3\right)}{\operatorname{\mathcal{D}}_{#1}(#2,#3)}{\operatorname{\mathcal{D}}_{#1}(#2,#3)}{\operatorname{\mathcal{D}}_{#1}(#2,#3)}}
  }
}
\newcommand{\fidelity}[3][]{
  \ifthenelse{\equal{#2}{}}
  {\ifthenelse{\equal{#3}{}}
    {\mathcal{F}_{#1}}{}
  }{
    \ifthenelse{\equal{#1}{}}
    {\mathchoice{\operatorname{\mathcal{F}}\left(#2,#3\right)}{\operatorname{\mathcal{F}}(#2,#3)}{\operatorname{\mathcal{F}}(#2,#3)}{\operatorname{\mathcal{F}}(#2,#3)}}
    {\mathchoice{\operatorname{\mathcal{F}}_{#1}\left(#2,#3\right)}{\operatorname{\mathcal{F}}_{#1}(#2,#3)}{\operatorname{\mathcal{F}}_{#1}(#2,#3)}{\operatorname{\mathcal{F}}_{#1}(#2,#3)}}
  }
}

\newcommand{\Sr}[3][]{
  \ifthenelse{\equal{#1}{}}
    {\operatorname{\mathnormal{S}}(#2\|#3)}
    {\operatorname{\mathnormal{S}}_{#1}(#2\|#3)}
}

\DeclareMathOperator{\1}{\mathds{1}}



\newcommand{\mb}[1]{\mathbb{#1}}

\newcommand{\Z}{\mb{Z}}

\definecolor{jens}{rgb}{0.1,0.5,0.1}
\definecolor{martin}{rgb}{0,0,1.0}

\newcommand{\beq}[0]{\begin{equation}}
\newcommand{\eeq}[0]{\end{equation}}

\newcommand{\hide}[1]{}

\begin{document}

\title{Fermionic topological quantum states as tensor networks}
\author{C. Wille, O. Buerschaper, and J. Eisert}
\address{Dahlem Center for Complex Quantum Systems, Freie Universit{\"a}t Berlin, 14195 Berlin, Germany}

\begin{abstract}
Tensor network states, and in particular projected entangled pair states, play an important role in the description of strongly correlated quantum lattice systems. They do not only serve as variational states in numerical simulation methods, but also provide a framework for classifying phases of quantum matter and capture notions of topological order in a stringent and rigorous language. The rapid development in this field for spin models and bosonic systems has not yet been mirrored by an analogous development for fermionic models. In this work, we introduce a framework of tensor networks having a fermionic component capable of capturing notions of topological order. At the heart of the formalism are axioms of fermionic matrix product operator injectivity, stable under concatenation. Building upon that, we formulate a Grassmann number tensor network ansatz for the ground state of fermionic twisted quantum double models. A specific focus is put on the paradigmatic example of the fermionic toric code. This work shows that the program of describing topologically ordered systems using tensor networks carries over to fermionic models.
\end{abstract}

\maketitle

One of the long-standing questions of theoretical physics  is concerned with the classification of \emph{phases of matter} in quantum many-body systems
\cite{WenBook}.
The issue at the heart of the matter is whether one can transform a given ground state of a local Hamiltonian 
into another ground state, without having to close the gap of the Hamiltonian along the path, in the presence or absence of symmetries.
In the simplest case of one spatial dimension, this question can be considered essentially 
settled. In the absence of symmetries, it turns out that there exists only a single phase. In the presence of 
symmetries, group cohomology provides a concise framework of labeling the different phases \cite{ClassificationPhases,WenPhases}.
In the solution of this problem, notions of \emph{tensor network states} \cite{OrusReview,VerstraeteBig,EisertReview,Schollwock201196}
provide the key, concepts that can be located between condensed-matter physics and quantum information theory. It is one of the important examples for showing how 
 ideas of \emph{entanglement theory} help making significant progress on questions in condensed matter theory.

Conceptually important as this question is, the situation at hand is still less clear beyond this comparably simple one-dimensional case. It is not quite 
settled yet what ``kinds of topological order'' one can find.
Progress on this difficult question has been made for spin models: Tensor network states, specifically \emph{projected entangled pair states} (PEPS) 
\cite{PEPSOld,PEPSKagome,iPEPS}, still provide a solid
machinery. 
Indeed, beyond their use as variational states in powerful numerical methods,
they provide a concise framework
for capturing notions of topological order in two spatial dimensions and beyond. 
Indeed, in the endeavour of classifying quantum phases of matter, it has been an extremely successful approach in the past to shift the emphasis slightly
and put the quantum state into the focus of attention, rather than the Hamiltonians. Quantum states can be
captured as tensor network states, in particular MPS in one spatial dimension and PEPS in higher dimensions; the link to the 
original Hamiltonian notion is made by means of the so-called \emph{parent Hamiltonian}, the Hamiltonian for which the 
tensor network states are the exact ground states. 

The classification and description of phases of matter and notions of topological order in terms of such 
projected entangled pair states has given rise to an enormously fruitful program for spin and bosonic systems. 
So-called injective PEPS exhibit a one-to-one correspondence between
the physical and virtual degrees of freedom \cite{PEPSTopology}. This implies that they are 
unique ground states of their parent Hamiltonians and not able to incorporate topological order yet. 
After all, the most obvious  topological invariant is a robust ground
state degeneracy which depends on the genus of the underlying surface. 
However, one can impose a  group symmetry and require that the
virtual and physical level of any local tensor region are
equivalent up to this group symmetry.  This is known as \emph{$G$-injectivity}~\cite{PEPSTopology} and leads to PEPS
describing the topological order of discrete gauge theories or
\emph{Kitaev's quantum double models} \cite{Kitaev2003fault} of which the famous \emph{toric
code} is an important example. $G$-injectivity, however, is not the only
mechanism which leads to topological order in
tensor networks. PEPS exhibiting \emph{matrix product operator injectivity} 
\cite{Buerschaper-AnnPhys-2014,Williamson2014,1409.2150}
go beyond this framework and are able to also represent \emph{string-net models}.
This is remarkable progress: 
But all this applies to spin models (or bosonic ones, for that matter). Fermionic topological order remains unaddressed in this program, however, to date. 

It is the purpose of this work to introduce a framework of tensor network states capable of capturing topological order in 
quantum lattice models having a fermionic component. In the focus of attention are from the start tensor network states
that exhibit a symmetry that can be formulated in terms of \emph{matrix-product operators}. We introduce \emph{axioms} for
\emph{matrix-product injectivity} that concisely capture the properties the involved tensors must have. Building upon this framework, we 
show that matrix-product operator symmetry is stable under concatenation. We connect the general approach to 
a large class of physical models by discussing fermionic twisted quantum double models. Much in the 
focus of attention is the specific example of the paradigmatic
\emph{fermionic toric code} \cite{FermionicToricCode}. 
In that, we build upon insights that have been obtained on fermionic symmetry protected topological order
\cite{GuWen}, as well as on frameworks to capture two-dimensional fermionic and bosonic topological order \cite{ChenGuWen2011,LinLevin,GaiottiKapustin}.
Fermionic tensor networks as such have been considered  \cite{MERAF2,MERAF3,MERAF1,Corboz2DHubbard,MERAF4,CorbozPEPSFermions},
and so have fermionic Gaussian PEPS \cite{WahlTuSchuchCirac2013,DubailRead}, 
ground states of non-interacting models, asking questions whether chirality could be achieved. The connection to notions of intrinsic topological order or 
matrix-product operator injectivity has not been achieved yet. It is the aim of this work to present a first significant step in this direction.

\begin{figure}
\centering
\includegraphics[width=0.18\textwidth]{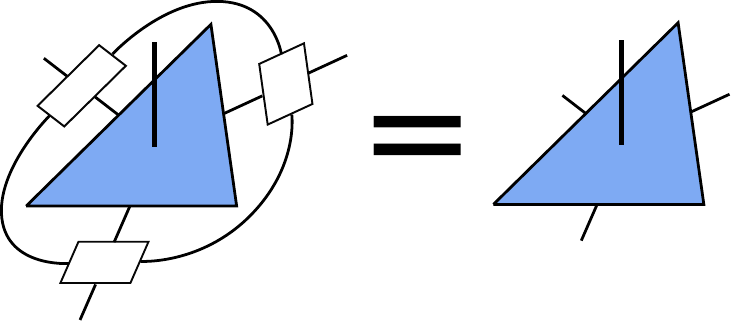}\,\,\,\,\,\includegraphics[width=0.18\textwidth]{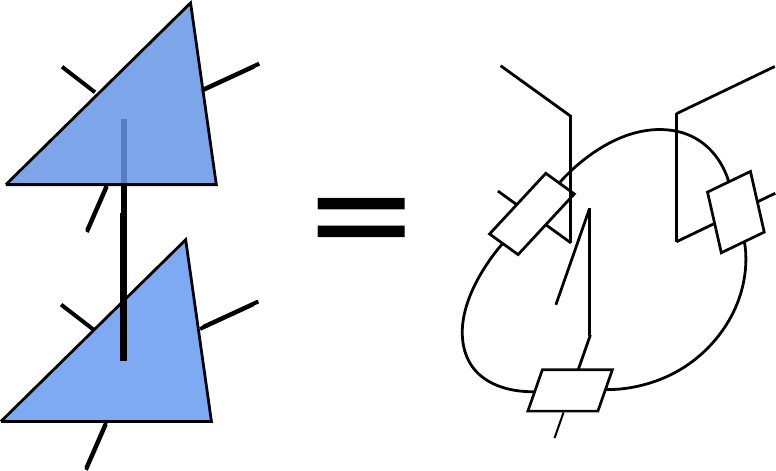}
\caption{MPO-symmetry (left) and MPO-injectivity (right).}
\label{mposym}\label{mpoinj}
\end{figure}

{\it Setting and notation.} We begin by defining the concepts  that will be relevant later on.
Let us first review the well-known notion of a bosonic PEPS \cite{PEPSOld,OrusReview,VerstraeteBig,EisertReview,Schollwock201196}: 
For a translation invariant system of $N$ spins (bosons) with local physical 
Hilbert space of dimension $d$ with basis states $\{\ket{i_j}\}_{i_j=1}^d$, 
a PEPS is a state vector defined by tensors $A$, equipped with an open physical index as well as virtual indices 
of dimension $D$ that are contracted with nearest neighbours in the lattice,
 via
\begin{equation}
\ket{\Psi} = \sum_{i_1,\ldots,i_N}   T \left[ A^{i_1} \ldots A^{i_N} \right] \ket{i_1,\ldots,i_N} \;, \label{PEPS}
\end{equation}
where $T$ denotes a contraction of all virtual indices that are connected according to a virtual lattice geometry. 
Each PEPS is the exact ground state of a respective \emph{parent Hamiltonian}. Virtual symmetries of the tensors $A$ characterised by 
\emph{matrix product operators} (MPO) can provide significant insights to the structure of the parent Hamiltonian. For the characterisation of intrinsic topological 
order the above mentioned notion of MPO-injectivity is of particular importance: 
A tensor is called \emph{MPO-injective} if it has a virtual symmetry given by a matrix product projector $P$ such that the tensor viewed as a map from the physical to the virtual space can be inverted on the virtual symmetry subspace given by $P$ (cf.\ Fig.~\ref{mposym}).  
The simplest instance of an MPO-injective PEPS is the \emph{famous toric code} \cite{Kitaev2003fault}, the most paradigmatic model of intrinsic topological order
and starting point for topological quantum memories. 

Yet again, the key step to the description of topological order of fermionic systems within the tensor network program is 
lacking entirely. This applies even to the paradigmatic {\it fermionic toric code} \cite{FermionicToricCode}. 
This model can be understood as the simplest \emph{fermionic string-net model} \cite{FermionicStringNet} -- a two-dimensional lattice model with spins on the edges and fermions on the vertices. Reminiscent of the bosonic 
analogue, its Hamiltonian
\begin{equation}
H= \sum_v Q_v + \sum_p Q_p \label{Hamiltonian}
\end{equation}
is given by a sum of vertex operators $Q_v$ and plaquette operators $Q_p$ that are mutually commuting local projectors acting on bosonic and fermionic degrees of freedom simultaneously. Even for such models, and more so in generality, 
significant obstacles remain. It goes without saying that a naive embedding into a spin system is doomed to failure 
as a consequence of Jordan-Wigner strings. The deviation from the bosonic case is also reflected by the fact that the low energy 
effective spin TQFT of the fermionic toric code model is given by a Chern-Simons theory that can not be realised by any local bosonic Hamiltonian. It is yet much 
 less clear to what extent ideas of MPO-injectivity potentially carry over, given that the notion of locality is largely altered by the presence of 
 fermionic anti-commutation relations. That is to say, the introduction of fresh 
concepts of fermionic tensor networks are necessary, which will be developed from now on.

{\it Fermionic tensor networks.} 
%
%
%
The key obstacle in constructing tensor networks for fermionic systems is to come up with a mechanism that allows the reordering of fermionic operators by keeping track of arising sign factors. Reordering the contraction order is a necessary prerequisite for the efficient contraction of a fPEPS. One possibility to solve this problem is to use PEPS of increased bond dimension instead of fPEPS \cite{SchuchFermiPEPS}. 
Another possibility is to use a particular kind of Grassmann number tensor network as e.g. in the fermion coherent state representation \cite{GuGrassmann}.

\begin{figure}
\centering
\large
\def\svgwidth{0.18\textwidth}
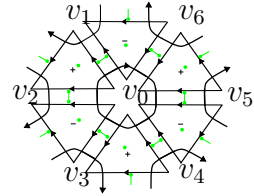
\caption{Fermionic projected entangled pair state (fPEPS). Green dots represent physical fermionic modes, green lines virtual fermonic bonds. Virtual bosonic bonds are denoted by black lines; bosonic physical indices at the edges of a triangle are not drawn explicitly.}
\label{hex}
\end{figure}

Here we use a similar construction. Each tensor has a physical fermionic mode $\theta^p$ and virtual fermionic modes $\theta^{f_i}$ represented by Grassmann numbers. In addition,  the tensor has bosonic physical and virtual indices $p_i$ and $v_i$ and a bosonic weight which is used to perform sign-factor book-keeping but also allows to describe fermion-spin hybrid systems. Defined on a trivalent lattice as drawn in Fig.\ \ref{hex} the tensors are of the form
\begin{align}
A= &\sum  A^{p_1p_2p_3 v_1v_2v_3}_{p f_1f_2f_3} \; \theta^p \theta^{f_1}  \bar \theta^{f_2} \bar \theta^{f_3} \ket{p_1,p_2,p_3}\bra{v_1,v_2,v_3} \;. \label{A_general}
\end{align} 
When the sum of all Grassmann number exponents is even, $\operatorname{Mod}[p+f_1+f_2+f_3,2]=0$, the tensors commute and thus the tensor network state is independent of the contraction order.

While it is clear what is meant by a contraction along a bosonic bond, a contraction along a Grassmann valued bond is defined by inserting a factor $\int \operatorname {d \bar \theta^f d \theta^f}$ and integrating out the two Grassmann numbers $\theta^f$ and $\bar \theta^f$.
Before the integration can be carried out a reordering of Grassmann numbers has to be performed yielding sign factors that depend on the Grassmann numbers of the tensors adjacent to the contracted fermionic bond. This property is relevant when one considers the virtual symmetries of a tensor. Fermionic symmetries captured by operators with fermionic bonds will always interfere with the bosonic symmetries as the contraction of fermionic bonds will yield additional sign-factors. Thus, the symmetries of the bosonic and fermionic degrees of freedom do not factorize unless the fermionic symmetry is trivial.

{\it Axioms of MPO-injectivity.} 
%
%
%
%
%
%
%
%
%
%
The key idea of fermionic MPO-injectivity remains the same as in the bosonic setting: The MPO projector singles out a virtual subspace on every region which eventually gives rise to the right entanglement scaling and the topological ground state degeneracy. 
However, the algebraic structure of the fermionic symmetry MPO and the corresponding virtual subspace is fundamentally different from the bosonic case. E.g. performing a Jordan-Wigner transform to obtain a usual bosonic MPO would drastically alter the locality structure of the MPO. The virtual fermions also affect the stability of MPO-symmetry and MPO-injectivity under concatenation, since concatenation of two MPOs is always accompanied by a reordering of virtual modes and thus by emerging additional sign factors. In the following, we will define fermionic matrix product operators and state the \emph{axioms of fermionic MPO-injectivity}:

\begin{definition}[fMPO-symmetry] 
A tensor $A$ has fermionic MPO-symmetry if it is invariant under the action of a fermionic MPO.
\end{definition}
Again, by fermionic MPO we refer to an MPO with virtual bosonic and fermionic degrees of freedom, and again
Grassmann numbers describe fermionic virtual modes. In order to formulate more general MPO-symmetries we have to introduce the concept of a 
\emph{branching structure}. That is all edges of a PEPS tensor are oriented in a way such that no cyclic orientation arises. Already in the bosonic case a large class of PEPS are symmetric under an MPO that depends on the {\it branching structure} \cite{Buerschaper-AnnPhys-2014}. The MPO itself is equipped with an orientation that we choose counter-clockwise as a matter of convention and consists of two different types of tensors: a $T_+$ tensor is located at edges that are parallel to the MPO direction and a $T_-$ tensor at anti-parallel edges.  

In the fermionic setting additional sign factors emerging from a reordering of fermionic modes pose an obstacle to the stability of MPO symmetry under concatenation -- an axiom that is required. In order to overcome this adversity and consistently define symmetry MPOs for a tensor of arbitrary size one can either increase the bond dimension of the respective $T_+$ and $T_-$ tensors or introduce an additional purely bosonic and diagonal sign-factor tensor $Y$ whose position within the MPO depends on the branching structure of the boundary. In the following we chose the latter approach.

\begin{axiom}[Projector property]
The two minimal meaningful MPOs defined for a triangular boundary with two different kinds of branchings structure are $P_+=\operatorname{tTr}[T_+ T_+  T_- Y]$ and $P_-=\operatorname{tTr}[T_+T_-T_-Y]$, where $\operatorname{tTr}$ denotes a tensor trace, i.e. a contraction of all bonds along the transversal direction of the MPO ring.
The MPOs fulfil 
%
$P_\pm^2=P_\pm$ and thus are projectors. \end{axiom}
Based on the stability under concatenation explained below this property generalizes to an MPO defined on a boundary of arbitrary size.

\begin{figure}
\centering
\includegraphics[width=0.35\textwidth]{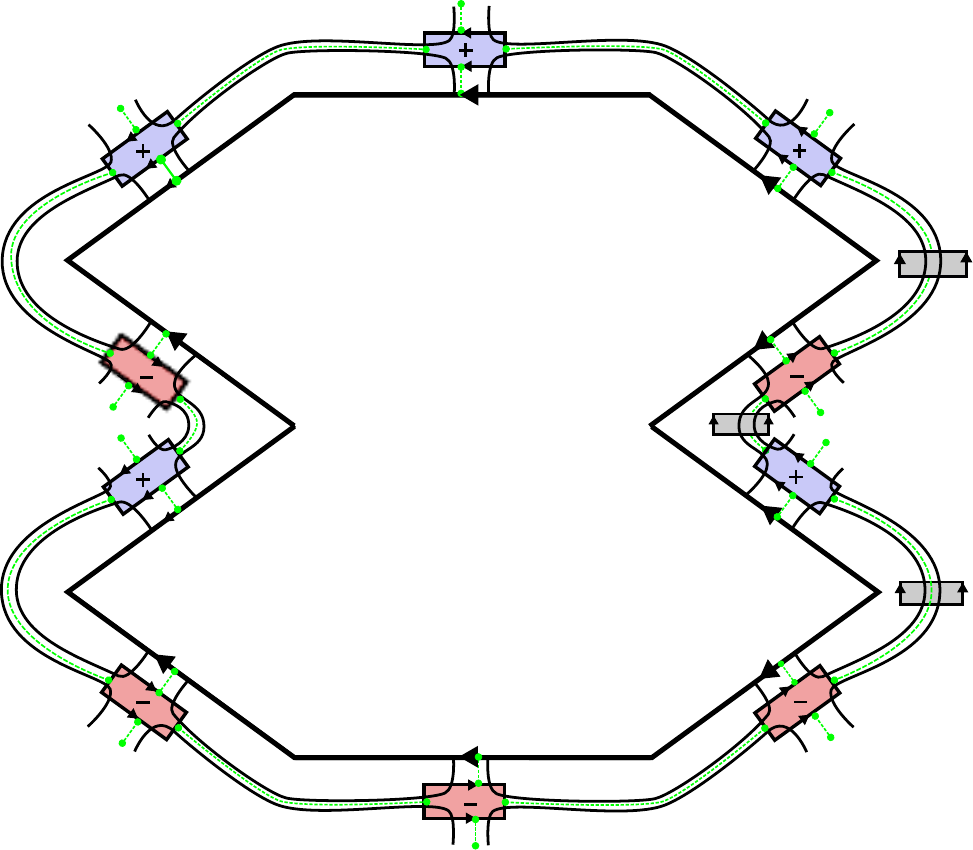}
\caption{MPO for a generic boundary consisting of  Y-tensors (grey) and tensors $T_+$ (blue) and $T_-$ (red). }
\label{boundarympo}
\end{figure}

\begin{axiom}[Stability of fMPO-symmetry]
Fermionic MPO-symmetry is stable under concatenation.
\end{axiom}
In particular, the concatenation of two fMPO-symmetric tensors with compatible branching structure is again fMPO-symmetric.
In order for this axiom to be realizable one first needs to consistently define symmetry fMPOs on regions of arbitrary size. One way to do this is to choose a branching structure of the tensor network that admits a global flow direction, i.e. all edges are oriented into one direction with a deviation of less than $\pi/2$ and to use a branching structure dependent placement of $Y$-tensors. In particular, $Y$-tensors are placed at each point of the boundary where the vertex matches one of the following criteria. 1) The boundary vertex has two outgoing edges which form an angle of less than $\pi$. 2) The boundary vertex has two incoming edges which form an angle of more than $\pi$. Here, the angle is defined as the angle inside the enclosed region. In Fig.\ \ref{boundarympo} we give an example of an MPO for a boundary branching structure that has vertices of either type and is representative for a generic MPO. 

\begin{figure}
\centering
\includegraphics[width=0.5\textwidth]{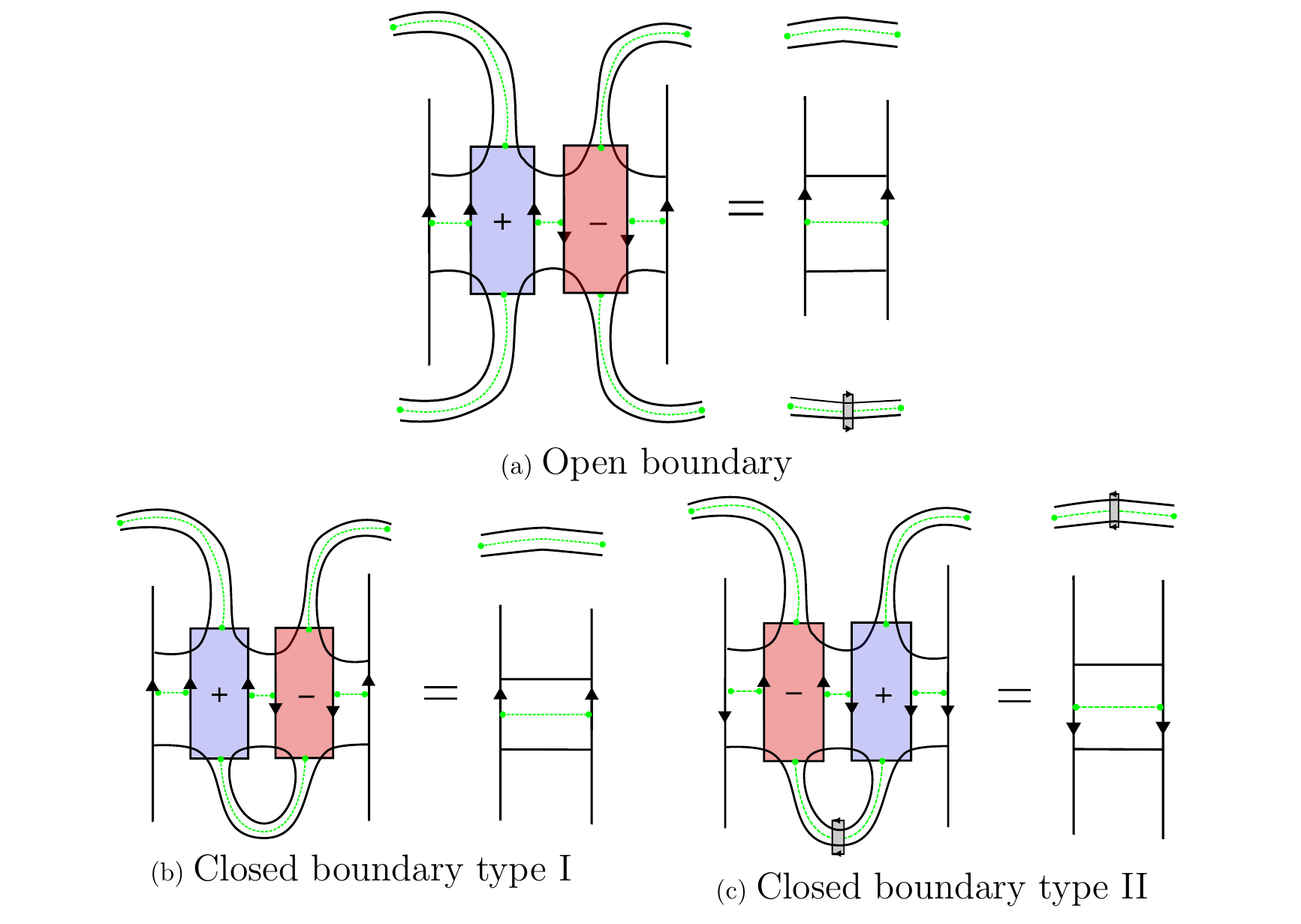}
\caption{Concatenation of two MPO tensors with open or partially closed radial indices.}
\label{concat}
\end{figure}

One can then show that if the concatenation of two MPO tensors proceeds according to the rules stated in Fig.\ \ref{concat} stability under concatenation is guaranteed. The proof is done by induction, showing that adding an MPO-symmetric triangle tensor to an existing MPO-symmetric tensor patch is again MPO-symmetric. Further details can be found in the appendix. 

%

\begin{definition}[fMPO-injectivity]
A tensor $A$ that is fMPO-symmetric and has a pseudo-inverse $\tilde A$ such that 
\begin{equation}
\tilde A A = P \;, \label{injective}
\end{equation}
i.e. is invertible on the fMPO-symmetric subspace given by the projector $P$ is called \emph{fMPO-injective}.
\end{definition}
\begin{axiom}[Stability of fMPO-injectivity]
Fermionic MPO-injectivity is stable under concatenation, i.e. the concatenation of two fMPO-injective tensors is again fMPO-injective. 
\end{axiom}
As in the bosonic case this is a direct consequence of the stability of fMPO-symmetry under concatenation \cite{Buerschaper-AnnPhys-2014}.

Tensor networks satisfying fermionic MPO-injectivity defined by the three axioms above provide a consistent and versatile
framework to describe non-chiral intrinsic topological order for fermionic models. The stability under concatenation is essential, 
but it is not trivial to see if and how it can be fulfilled. To provide further substance to the framework established, we
present a large class of fermionic models that admit an fMPO-injective tensor network description. 

{\it Fermionic twisted quantum double models.}
In the realm of bosonic MPO-injectivity it has been established that the ground states of an important class of models exhibiting topological order, i.e. Levin-Wen string net models can be written as PEPS. Furthermore, the self-consistency equations of the renormalization group (RG) flow defined for Levin-Wen string-net models, in particular the pentagon equation give rise to a virtual MPO-symmetry of the respective tensors. 

Building upon this insight we construct fPEPS for a particular subset of fermionic string-net models proposed in Ref.~\cite{FermionicStringNet} that we refer to as {\it fermionic twisted quantum double models}. Here, the bosonic spin degrees of freedom (edge-labels) are given by a group $G$ and spinless fermions on the vertices are coupled to the bosonic edge degrees via a 2-cocyle $s \in \mathcal H^2(G,\mathrm \Z_2)$ fulfilling
\begin{equation}
\operatorname{Mod}[s(a,b)+ s(ab,c)+s(a,bc) +s(b,c),2]=0 \;. \label{2cocycleEq}
\end{equation}
I.e. the presence or absence of a fermion  at a vertex is determined by the adjacent spins on the two incoming edges or the two outgoing edges $g_i,g_j$ of the vertex via $\theta^{s(g_i,g_j)}$. 

For bosonic twisted quantum double models the main self-consistency equation of the RG flow, i.e., the pentagon equation reduces to a more simple form: a 3-cocycle equation. In the fermionic setting the anticommutation relations lead to a super-3-cocycle equation, i.e. a 3-cocycle equation graded by the 2-cocycle $s$
\begin{align}
&\omega(a,b,c) \omega(a,bc,d) \omega(b,c,d) \nonumber \\
 =& (-1)^{s(a,b)s(c,d)} \omega(ab,c,d) \omega(a,b,cd) \;. \label{gpentagon}
\end{align}
A solution to Eq.~(\ref{gpentagon}) exists iff the function $(-1)^{s(a,b)s(c,d)}$ is a coboundary $\mathcal B^4(G,U(1))$ viewed as a 4-cocycle. The full model is then described in terms of a triple $(G,s \in \mathcal H^2(G,\mathrm \Z_2), \omega \in \mathcal H^3_f(G,U(1),s))$, where $\omega \in \mathcal H^3_f(G,U(1),s)$ means that $\omega$ fulfils Eq.~(\ref{gpentagon}). 

\begin{figure}
\centering
\def\svgwidth{0.35\textwidth}
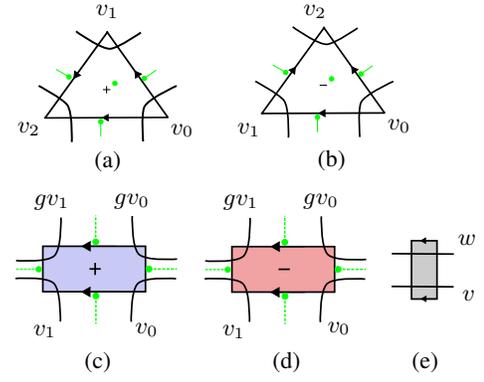
\caption{Tensors (a)$A_+$, (b)$A_-$, (c)$T_+(g)$, (d)$T_-(g)$ and (e)$Y$.}
\label{fig_legend}
\end{figure}

Motivated by the duality of MPO-injectivity of PEPS and the self-consistency equations of the RG flow in string-net models, we construct the fPEPS tensors. For reasons that will become clear later, we use the branching structure of the lattice to define positively and negatively oriented tensors, i.e. a triangle tensor $A_{+/-}$ has positive (negative) orientation, if the majority of its edges is oriented clockwise (counter-clockwise). 
The expressions for $A_+$ and $A_-$ depicted in Fig. \ref{fig_legend}(a) and (b) read
\begin{align}
A_+ =& \sum_{v_0,v_1,v_2} \omega(v_0, v_0^{-1}v_1, v_1^{-1}v_2) \label{Afinal} \\
\times &\theta^{s( v_0^{-1}v_1, v_1^{-1}v_2)} \theta^{s(v_0, v_0^{-1}v_2)}  \bar \theta^{s(v_1, v_1^{-1}v_2)}  \bar  \theta^{s(v_0, v_0^{-1}v_1)} \nonumber\\
\times&  \ket{ v_0^{-1}v_1, v_1^{-1}v_2, v_0^{-1}v_2}\bra{v_0,v_1,v_2} ,\nonumber\\
A_- =& \sum_{v_0,v_1,v_2} \omega^{-1}(v_0, v_0^{-1}v_1, v_1^{-1}v_2) \label{A-final} \\
\times&\theta^{s(v_0, v_0^{-1}v_1)} \theta^{s(v_1, v_1^{-1}v_2)}  \bar  \theta^{s(v_0, v_0^{-1}v_2)} \bar \theta^{s( v_0^{-1}v_1, v_1^{-1}v_2)} \nonumber \\
\times&  \ket{v_0^{-1}v_1, v_1^{-1}v_2, v_0^{-1}v_2}\bra{v_0,v_1,v_2 } \nonumber  \;.
\end{align}
The 2-cocycle $s$ that defines the presence of the physical fermion is also used to determine the presence of virtual fermions via a coupling to their adjacent virtual spins. 
This construction yields commuting fPEPS tensors due to the 2-cocycle property of $s$ given in Eq.~(\ref{2cocycleEq}). Reminiscent of the bosonic setting the weight of the tensors is defined in terms of the super-3-cocycle $\omega$ and the relation between physical and virtual indices is chosen as $p_{i,j}=v_i^{-1}v_j$, which means that the fPEPS is not injective, but allows for injectivity on a symmetric subspace.

While the basic construction of the fPEPS can be motivated from a string-net picture the details of the formalism rely on the established connection between MPO-injective PEPS and Dijkgraaf-Witten discrete path integral \cite{Buerschaper-AnnPhys-2014}.
In particular, in the discrete fermionic path integral formalism describing fermionic symmetry protected topological order \cite{GuWen} a consistent assignment of orientations to simplices and faces is used to also consistently assign $\theta$ and $\bar \theta$ in a particular order to the respective simplices. We employ this consistent formalism in a modified way to construct the Grassmann tensor network for phases with intrinsic topological order.
The tensors defined in this way fulfill the axioms of fMPO-injectivity. To illustrate this, we restrict our analysis to a branching structure admitting a global flow and explicitly construct the $T_+$ and $T_-$ tensors of the fMPO depicted in Fig. \ref{fig_legend}(c) and (d)
\begin{align}
T_+(g)=&\sum_{v_0,v_1} \omega(g,v_0, v_0^{-1}v_1)   \quad \ket{v_0,v_1 }\bra{g v_0 ,g v_1} \label{T+}\\
\times&\theta^{s( v_0,  v_0^{-1} v_1)} \theta^{s(g , v_1)} \bar \theta^{s(g v_0,  v_0^{-1}v_1)} \bar \theta^{s(g,v_0)} ,  \nonumber \\
T_-(g)=&\sum_{v_0,v_1} \omega^{-1}(g,v_1, v_1^{-1}v_0)   \quad \ket{v_0,v_1 }\bra{g v_0 ,g v_1} \label{T-}\\
\times&\theta^{s(g,v_1)} \theta^{s(g v_1,  v_1^{-1}v_0)} \bar \theta^{s(g , v_0)}   \bar \theta^{s( v_1,  v_1^{-1} v_0)}  \nonumber
\end{align}
and the purely bosonic sign factor tensor depicted in Fig. \ref{fig_legend}(d)
\begin{equation}
Y = \sum_{v,w} (-1)^{s(w v^{-1},v)} \ket{v,w}\bra{v,w} \label{Ytensor} \;,
\end{equation} 
where each group element $g \in G$ yields a distinct symmetry MPO, i.e. we have
\begin{equation}
A_\pm = A_\pm V_\pm(g) \;, \label{symmetry}
\end{equation}
with $V_\pm(g)=\text{tTr}[T_\pm(g)T_\pm(g) T_\mp(g) Y]$ for each element independently. The MPO projector is given by $P_\pm=\sum_g V_\pm(g)$ and the projector identity $P_\pm^2=P_\pm$ follows from the fact that $V_\pm(g)$ fulfils the group representation property  $V_\pm(g) V_\pm(h)=V_\pm(hg)$.
The pseudo-inverses of $A_\pm$ can be calculated explicitly and are given by
\begin{align}
\tilde A_+ =& \sum_{v_0,v_1,v_2} \omega^{-1}(v_0, v_0^{-1}v_1, v_1^{-1}v_2) (-1)^{s(v_0, v_0^{-1}v_2)} \nonumber\\
\times& \theta^{s(v_0, v_0^{-1}v_1)} \theta^{s(v_1, v_1^{-1}v_2)} \bar \theta^{s(v_0, v_0^{-1}v_2)} \bar \theta^{s( v_0^{-1}v_1, v_1^{-1}v_2)} \nonumber \\
\times&  \ket{v_0,v_1,v_2}\bra{ v_1^{-1}v_2, v_0^{-1}v_1, v_0^{-1}v_2} , \label{Atilde}\\
\tilde A_- =& \sum_{v_0,v_1,v_2} \omega(v_0, v_0^{-1}v_1, v_1^{-1}v_2) (-1)^{s(v_0, v_0^{-1}v_2)} \nonumber\\
\times&\theta^{s( v_0^{-1}v_1, v_1^{-1}v_2)} \theta^{s(v_0, v_0^{-1}v_2)}  \bar \theta^{s(v_1, v_1^{-1}v_2)}  \bar  \theta^{s(v_0, v_0^{-1}v_1)} \nonumber \\
\times&  \ket{v_0,v_1,v_2}\bra{ v_1^{-1}v_2, v_0^{-1}v_1, v_0^{-1}v_2} \;. \label{A-tilde}
\end{align}
The stability of MPO-symmetry and MPO-injectivity under concatenation follows directly from the fact that the tensors $T_\pm$ fulfil the concatenation properties depicted in Fig.\ \ref{concat} which is a consequence of direct calculation.

{\it Ground state space.}
The concept of fMPO-injectivity makes it possible to compute essential properties of a physical model solely based on the characterization of the virtual tensor symmetries without considering the state or the Hamiltonian explicitly on a physical level. In the following we compute the ground state degeneracy of an fMPO-injective PEPS on a torus equipped with a branching structure admitting a global flow. We closely follow the approach 
of Ref.~\cite{Buerschaper-AnnPhys-2014}.

\begin{figure}
\centering
\includegraphics[width=0.45\textwidth]{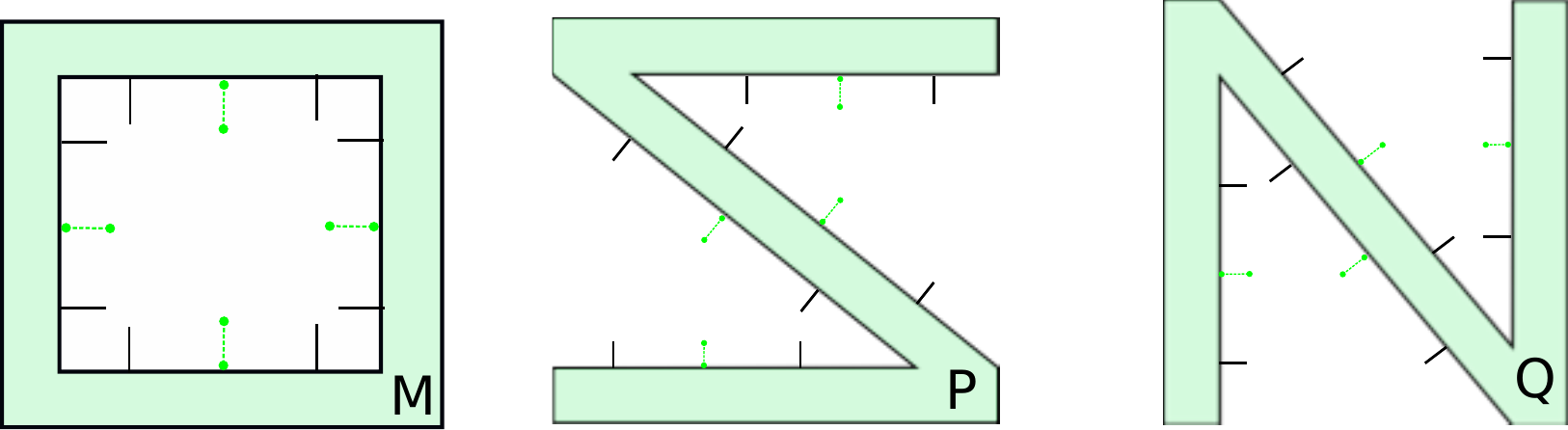}
\caption{The three different closure tensors on a minimal torus.}
\label{closuretensors}
\end{figure}

\begin{figure}
\centering
%
\includegraphics[width=0.45\textwidth]{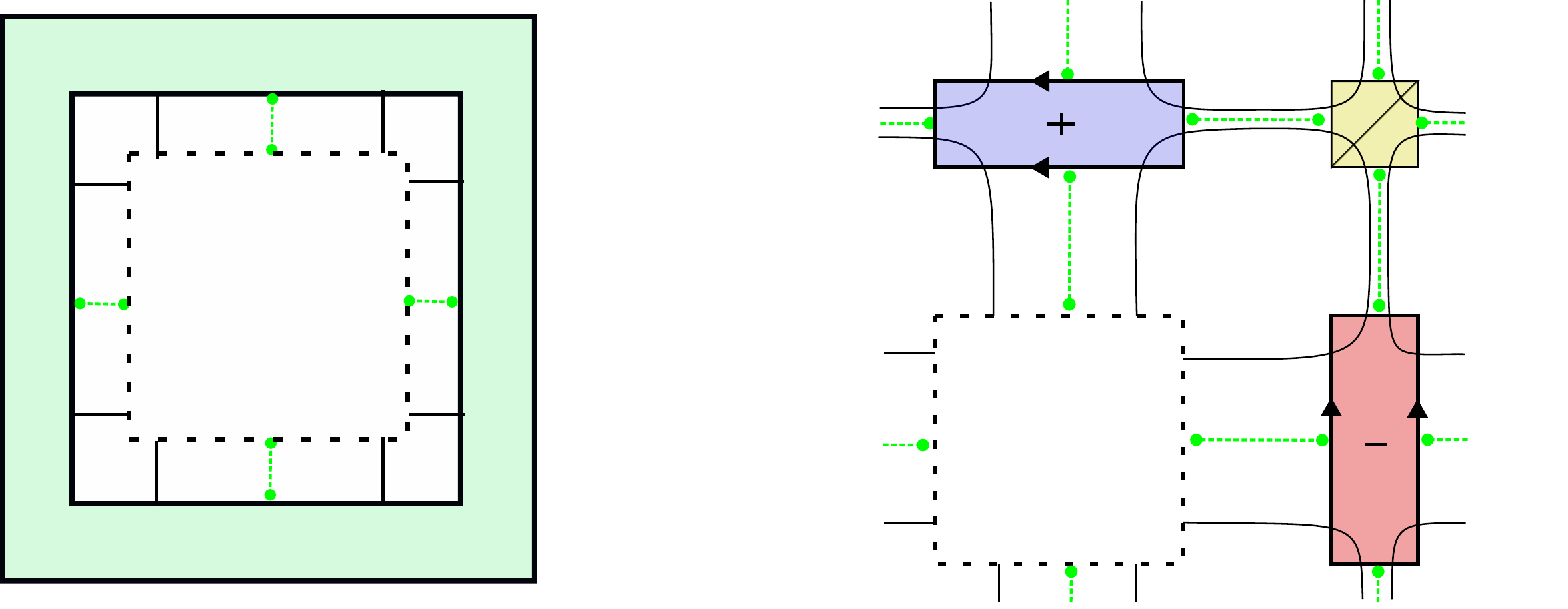}
\caption{Closure tensor $M(g,h)$.}
\label{Mtensor}
\end{figure}

We consider all locally undetectable closures on a minimal torus which are MPO-symmetric, i.e. all states which can be defined simultaneously using the three different closure tensors depicted in Fig.\ \ref{closuretensors}. Similar to the bosonic setting one can parametrize this space as
\begin{equation}
\text{span}\{M(g,h) | \;[g,h]=0,\; s(g,h)=s(h,g) \} \;,
\end{equation}
where the closure tensor $M$ is constructed as depicted in Fig.~\ref{Mtensor} and given by
\begin{align}
M(g,h) =& \sum_\alpha \lambda(\alpha;g,h)  \ket{\alpha,g \alpha , gh \alpha , h \alpha } \label{Mgh} \\
&\times \theta^{s(g\alpha,^\alpha g^{-1}) } \theta^{s(hg\alpha,^\alpha  h^{-1} )} \bar \theta^{s(g h \alpha, ^\alpha g^{-1} )} \bar \theta^{s(h \alpha, ^\alpha h^{-1})}\nonumber   \;.
\end{align}
Here, $^\alpha g=\alpha^{-1}g\alpha$ denotes conjugation by the inverse element. The full expression for $\lambda(\alpha;g,h)$ is given in the appendix.

Imposing MPO-symmetry simply by acting with a four-site MPO $V(k)$ on $M$ we obtain a parametrization of the ground state space in terms of MPO-symmetric tensors
\begin{equation}
M'(g,h)=\frac{1}{|G|} \sum_{k \in G} V(k) M(g,h) \;. \label{Mprimegh}
\end{equation}
Next we count the number of linearly independent elements of the ground state space. First, one can show that two states $M'(g,h)$ and $M'(k,l)$ are linearly dependent if $(g,h)$ and $(k,l)$ are in the same pair conjugacy class. Furthermore, not all pair conjugacy classes correspond to a non-vanishing state. As in the bosonic case one finds that only $c$-regular pair conjugacy classes contribute to the ground state dimension. Defining
\begin{equation}
c_g(h,k)=\frac{\omega(g,h,k)\omega(h,k,g)}{\omega(h,g,k)} \label{comega}
\end{equation}
a pair conjugacy class $\mathcal C(g,h)$ is called $c$-regular if for all elements of the centralizer $k \in \mathcal Z(g,h)$ we have 
\begin{equation}
c_g(h,k)=c_g(k,h) \;.
\end{equation}
Details of the derivation are given in the appendix. 

Since states from different pair conjugacy classes are linearly independent we conclude that the ground state degeneracy on a minimal torus is given by the number of $c$-regular pair conjugacy classes $\mathcal C(g,h)$ for which $[g,h]=0$ and $s(g,h)=s(h,g)$ is fulfilled. In the case where $s=0$, i.e. the bosonic case, we obtain a classification of the ground state basis elements in terms of $c^\omega$-regular pair conjugacy classes as expected from the results of Ref.~\cite{Buerschaper-AnnPhys-2014}.
In complete analogy to the bosonic setting the considerations on the minimal torus hold equally well on a torus of arbitrary size due to the axioms of stability under concatenation.

{\it Fermionic toric code.}
To illustrate the formalism of fermionic MPO-injective fPEPS we will elucidate how the ground state of the fermionic toric code Hamiltonian as proposed in Ref.~\cite{FermionicToricCode} can be written as a fermionic tensor network satisfying the axioms of fermionic MPO-injectivity. The fermionic toric code Hamiltonian is given in a string-net description. 
It can be seen as the simplest solution of the self-consistency equations for fermionic twisted quantum double models, i.e., the simplest triple $(G,s,\omega)$.

The local Hilbert space is given by qubits represented by the group $\mathbb \Z_2=\{0,1\}$. This group has only two second cohomology classes. Choosing the 2-cocycle $s$ as a representative of the trivial cohomology class one obtains the usual bosonic toric code or the double semion model depending on the choice of $\omega$, i.e., the solution to the pentagon equation which is a usual 3-cocyclce equation in this case. Choosing the non-trivial second cohomology class, in particular the normalized representative
\begin{equation}
s(g,h)=\begin{cases} 1  \quad & \text{ if } g=h=1\;, \\ 0 \; & \text{otherwise} \end{cases} \label{ftc2cocycle}
\end{equation}
corresponds to the rule according to which fermionic creation and annihilation takes place in the fermionc toric code model. The two normalized solutions to the graded pentagon equation
\begin{equation}
\omega(g,h,k) = \begin{cases} \pm \mathrm i  \quad & \text{ if } g=h=k=1\;, \\ 1 \; & \text{otherwise} \end{cases} \label{ftcg3cocycle}
\end{equation}
then correspond to the fermionic toric code model or its dual defined by choosing the fusion coefficient $\alpha =\pm \mathrm i$ in Ref.~\cite{FermionicToricCode}.

The ground state wave function of the fermionic toric code can be written in terms of $A_\pm$ tensors defined in Eq.~(\ref{Afinal}) and (\ref{A-final}) using $s$ and $\omega$ defined in Eq.~(\ref{ftc2cocycle}), and (\ref{ftcg3cocycle}) with the lattice geometry depicted in Fig.~\ref{hex} which is the dual of the physical lattice from Ref.~\cite{FermionicToricCode}. By construction this tensor network is fMPO-injective. To prove that it is indeed the ground state of the fermionic toric code Hamiltonian one can verify that the tensor network is an eigenstate to each local projector. The property of being an eigenstate to the vertex projector is also already implied in the construction of the $A_\pm$ tensor. To check that the tensor network state is an eigenstate of the plaquette projectors we compute a local patch $A_\text{hex}$ of the tensor network given in Fig.~\ref{hex} with open virtual boundary and act with the plaquette operator $Q_p$ on the physical indices. One can use Table~\ref{table_hamiltonian} stating the matrix elements of $Q_p$ explicitly in a suitable gauge to check that 
\begin{equation}
Q_p A_\text{hex}=A_\text{hex} \;. \label{plaquette}
\end{equation}
Intuitively this can be understood by interpreting the plaquette operator as inserting a closed loop around the center of the hexagon. The contraction of the inner virtual index $v_0$ exactly compensates for that, i,e., $Q_p A_\text{hex}(v_0=0)=A_\text{hex}(v_0=1)$ and $Q_p A_\text{hex}(v_0=1)=A_\text{hex}(v_0=0)$. Note that due to the anti-commutating Grassmann variables it is not trivial to see that all sign-factors are indeed correct but it is a result of an explicit calculation. Making use of the translation invariance of the lattice we have successfully verified that the tensor network state is the groundstate on any local region.

To calculate the ground-state degeneracy on a torus we use the formalism developed in the previous section. Since $\mathbb \Z_2$ is Abelian there are four pair conjugacy classes. As the cocycle $s$ is symmetric the condition $s(g,h)=s(h,g)$ is fulfilled as well. Next we check whether
\begin{equation}
\frac{c_g(h,k)}{c_g(k,h)}=1 \;, \label{cregular}
\end{equation}
which is true for all $g,h,k$. This is easily seen by noting that the ratio is one if any of the elements $g,h,k$ is equal to $0$ and if $g=h=k=1$ the ratio is also one, because the supercocycle $\omega$ is symmetric under permutation of the order of its arguments. Therefore, each pair conjugacy class spans one ground-state dimension and the ground-state degeneracy on a torus is four which is in agreement with the result obtained directly on the physical level in Ref.~\cite{FermionicToricCode}.

{\it Summary and outlook.} In this work, we have introduced a general tensor network formalism that is sufficiently
versatile to capture topological order of quantum systems with a fermionic component.
We hence generalise the idea of describing phases of matter using
tensor network states to the fermionic realm. Yet, this is only the beginning of an extensive program, needless
to say: In future work, instances of symmetry-protected topological order will be discussed, as well
as more subtle situations in which a global flow cannot be identified
\cite{LongFermionicTopological}. It will also be interesting to grasp modular matrices directly
in this framework. It is the hope that this work can be seen as a further invitation 
to explore tensor networks to capture topological phases of matter.

{\it Acknowledgements.} We would like to thank the  DFG (CRC 183), the 
Templeton Foundation, and the ERC (TAQ) for support.


%

\section{Appendix}

In this appendix, we provide further detail to the arguments of the main text, on the stability of fMPO-symmetry,
the identification of the ground state space, and the transformation of Hamiltonians under gauge transformations.

\subsection{Stability of fMPO-symmetry}

One can easily verify that fMPO-symmetry is stable under concatenation for a branching structure admitting a global flow by considering all different concatenation cases. To this end we first make a distinction between an ``open'' concatenation where the MPO tensor concatenation depicted in Fig.~\ref{concat}a is relevant and a partially closed concatenation (Fig.~\ref{concat}b,c) relevant when an outer vertex of a triangle tensor becomes an inner vertex during the concatenation process. 

\begin{figure}
\centering
\includegraphics[width=0.45\textwidth]{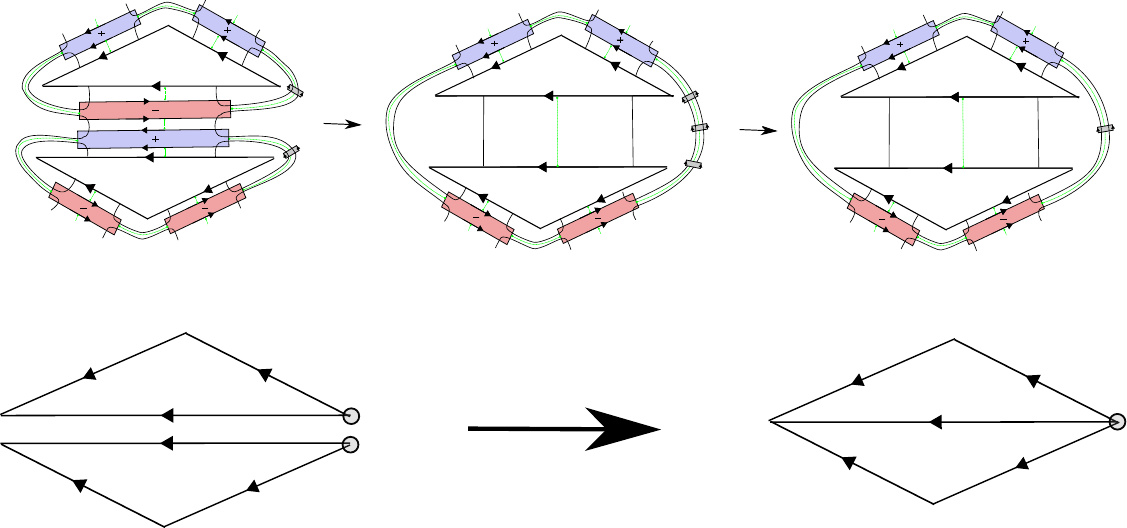}
\caption{The concatenation of two MPO-symmetric tensors written in explicit notation (upper panel) and in abbreviated notation only depicting the positions of $Y$-tensors. }
\label{abb}
\end{figure}

Before we discuss the different cases occurring 
we introduce a short-hand notation to symbolize the concatenation of MPO-symmetric tensors. Instead of drawing the full symmetry MPO, 
we just indicate the positions of $Y$-tensors before and after the concatenation (Fig.~\ref{abb}) by circles at the respective boundary vertices.

\begin{figure}
\centering
\includegraphics[width=0.4\textwidth]{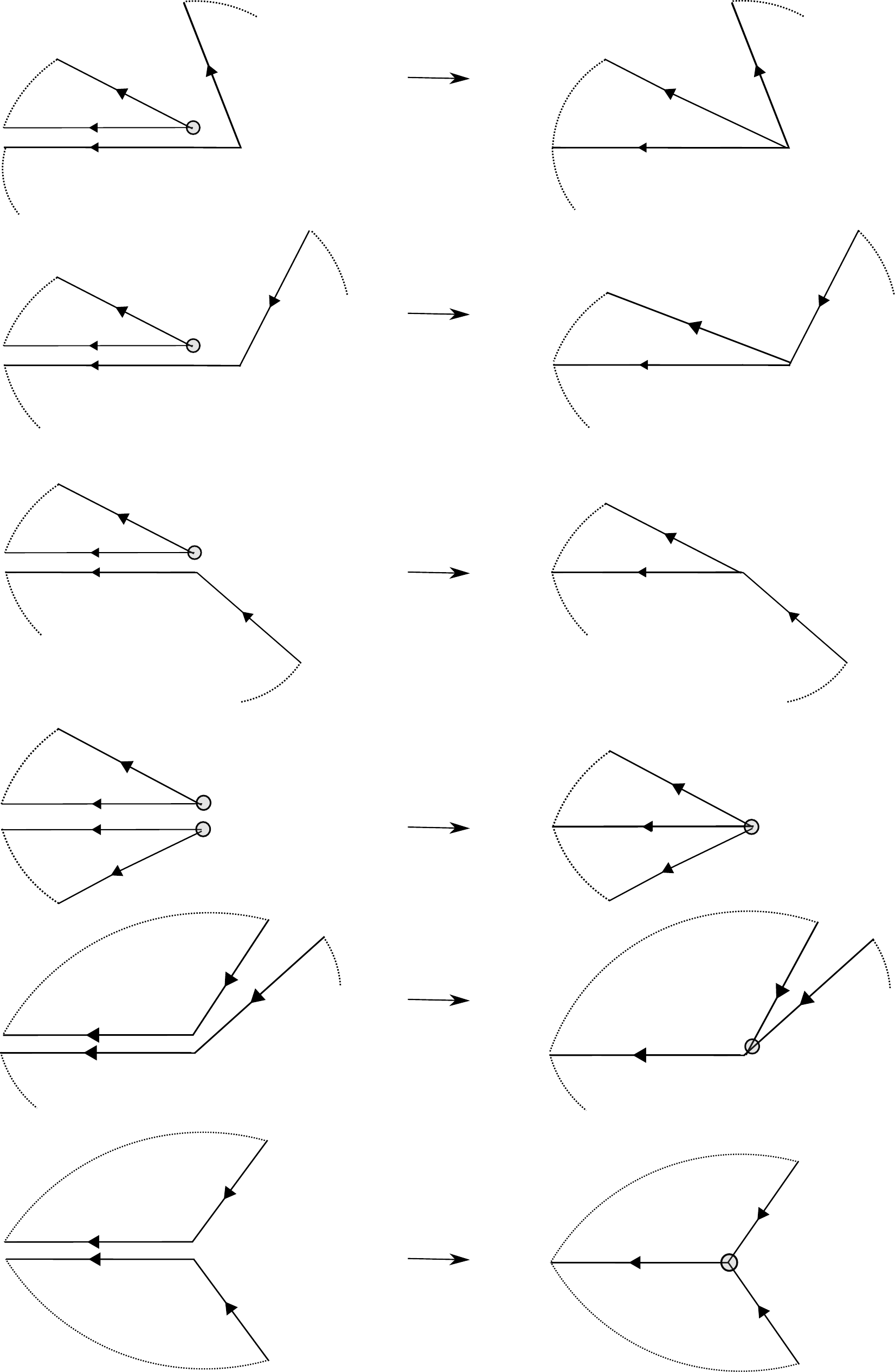}
\caption{All six distinct cases for concatenating two MPO-symmetric tensors along an open edge.}
\label{casesopen}
\end{figure}

In the case of ``open'' concatenation it is sufficient to consider concatenation along an edge pointing towards the global flow direction and then categorize the different cases according to the shared angles with the neighboring edges at the origin vertex. Since the placement of $Y$-tensors only depends on the the criteria whether angles are smaller or larger than $\pi$ and the edge orientations the exact angles are irrelevant and we only have to distinguish four cases distinguished by $0,\pi/2,\pi,3\pi/2$. Taking into account geometric constraints imposed by the global flow criterion and making use of the mirror symmetry along the global flow direction there are 6 distinct cases left to consider. All of them can be shown to yield the correct MPO-symmetry after concatenation as shown in Fig.~\ref{casesopen}. 

The second case is relevant when two MPO-symmetric tensors are contracted along a common boundary of length two or more. Performing the contraction sequentially along the common boundary amounts to a step-by-step reduction of the size of the boundary and thus to a reduction of the size of the symmetry MPO. Note that in this case there are two different concatenation rules depending on the orientation of the inner edge relative to the position of the already contracted transversal indices (Fig.~\ref{concat}b and c). 

\begin{figure}
\centering
\includegraphics[width=0.4\textwidth]{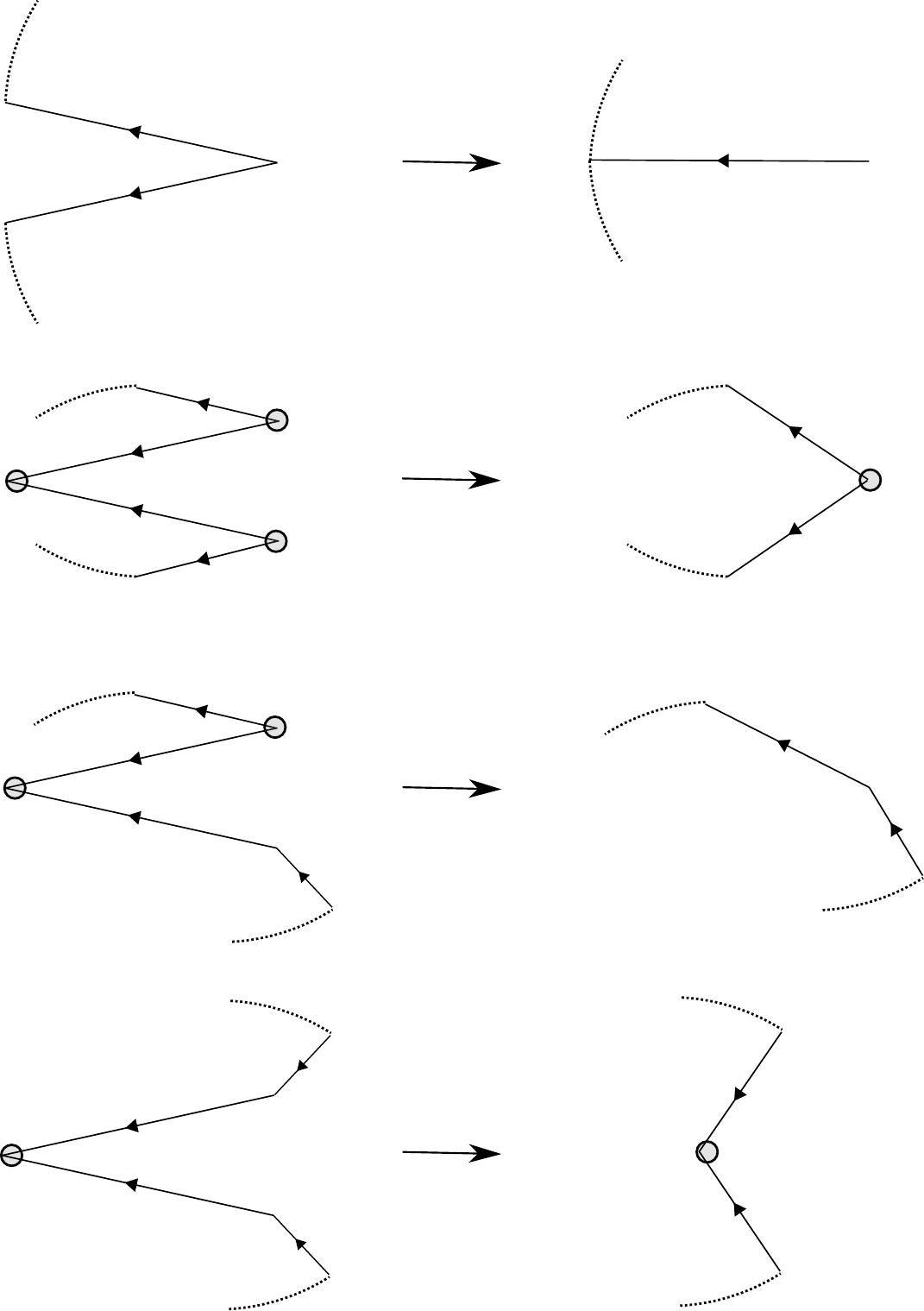}
\caption{Contracting two edges of the same MPO-symmetric tensor.}
\label{closingcases}
\end{figure}

If the edges to be contracted share a vertex at their origin the contraction does not yield additional $Y$-tensors and independently from the rest of the tensor the contracted tensor will have the expected MPO-symmetry (Fig.~\ref{closingcases} upper panel). In the other case, one can consider all different cases of the edge configurations in the immediate vicinity of the contracted edge. Using symmetry arguments, there are three distinct cases which can be checked explicitly (Fig.~\ref{closingcases} panel 2 to 4).

\subsection{ Ground state space} 
The coefficients $\lambda(\alpha;g,h)$ defining the state $M(g,h)$ in Eq.~(\ref{Mgh}) are given by
\begin{align}
\lambda(\alpha;g,h)=&\frac{\omega(h,g,\alpha) \omega(h, g\alpha, ^\alpha g^{-1} )}{\omega(g,h,\alpha) \omega(g,h\alpha ,^\alpha h^{-1} )} \label{lambda} \\
\times&(-1)^{s(gh \alpha,^\alpha h^{-1} ) s(g\alpha,^\alpha g^{-1} )+s(hg\alpha,^\alpha g^{-1}) s(h \alpha,^\alpha h^{-1} )} \nonumber \\
\times&(-1)^{(s(h,\alpha)+s(h\alpha,^\alpha h^{-1} ))(s(g,\alpha)+s(g\alpha,^\alpha g^{-1} ))+s(hg,\alpha)}. \nonumber  
\end{align}
The MPO-symmetric boundary state is given by
\begin{align}
M'(g,h) =& \frac{1}{|G|} \sum_k V(k) M(g,h) \label{Mprimeghappendix}\\
 =& \frac{1}{|G|} \sum_k \eta_g(h,k) M(g^k,h^k) \;, \nonumber 
\end{align}
where $g^k=kgk^{-1}$ denotes conjugation and
\begin{align}
\eta_g(h,k) =& \frac{\omega(g,k^{-1},h^k) \omega(k^{-1},h^k,g^k) \omega(h,g,k^{-1})}{\omega(h,k^{-1},g^k) \omega(k^{-1},g^k,h^k) \omega(g,h,k^{-1})} \nonumber \\
&(-1)^{[s(k^{-1},kh)+s(kh,k^{-1})][s(k^{-1},kg)+s(kg,k^{-1})]} \nonumber \\
&(-1)^{s(k^{-1},kgh)+s(kgh,k^{-1})} \label{eta} \;.
\end{align}
The fact that $M'(g,h)$ and $M'(j,l)$ are linear dependent if $(j,l)=(g^t,h^t)$ for some $t$, i.e. if they are in the same pair conjugacy class follows from the identity
\begin{equation}
\eta_{g^t}(x^t,yt^{-1}) = \frac{\eta_g(x,y)}{\eta_g(x,t)} \label{lemma10}
\end{equation}
that holds formally as in the bosonic setting despite the fact that $\eta$ has additional sign factors and is given by a product of super 3-cocycles $\omega$. Eq.~(\ref{lemma10}) is also used in order to derive that only $c$-regular pair conjugacy classes contribute to the ground state dimension. To this end, first note that if
\begin{equation}
\sum_{s\in Z(g,h)} \eta_g(h,s)=0 \;, \label{centralizer_coeff}
\end{equation}
then also
\begin{equation}
\sum_{s | g^s=g^{t_i}, h^s=h^{t_i}} \eta_g(h,s) =0 , \quad \forall i \;. \label{coeff}
\end{equation}
Writing out the state $M'(g,h)$ in the basis of the elements of the pair conjugacy class $\mathcal C(g,h)$ we make use of the fact above to conclude that
\begin{equation}
M'(g,h)=0 \Leftrightarrow \sum_{s\in\mathcal Z(g,h)} \eta_g(h,s)=0  \;.\label{coeff2}
\end{equation}
To single out the pair conjugacy classes for which Eq.~\ref{coeff2} is fulfilled we note that for all elements $k$ in the centralizer of $(g,h)$ with $[g,h]=0$ we have
\begin{equation}
 \eta_g(h,k) = \frac{c_g(k^{-1},h)}{c_g(h,k^{-1})} \;, \label{etac}
\end{equation}
where $c_g(h,k)$ as defined in Eq.~(\ref{comega}). For $g,h,k \in \mathcal Z(g,h)$ and $[g,h]=0$, i.e. $g,h,k$ are mutually commuting $c_g(h,k)$ is a 2-cocycle. This insight is used to apply the same arguments as in Ref.~\cite{TwistedQMDouble} and show that 
\begin{equation}
\sum_{s\in\mathcal Z(g,h)} \eta_g(h,s)=0  \Leftrightarrow c_g(h,s) \neq c_g(s,h) \; . \label{coeff3}
\end{equation}
In other words, only pair conjugacy classes $\mathcal C(g,h)$ contribute to the total ground state space dimension for which $[g,h]=0, s(g,h)=s(h,g)$ and $c_g(h,k)=c_g(k,h)$.

\subsection{Hamiltonian}
As addressed in Ref.~\cite{LinLevin} string-net models can be defined using different gauges. Under such gauge transformations the Hamiltonian changes according to local physical unitary operations that do not alter the topological phase. A typical gauge degree of freedom is the choice of the loop weight, also referred to as quantum dimension $d_i$. For all bosonic twisted quantum double models an MPO-injective PEPS description can be found. Here, a particular gauge degree of freedom is the choice of the representative $\omega$ of a certain cohomology class which defines the model. A natural gauge is to choose a normalized 3-cocycle, which will yield a wave-function invariant under adding a closed-loop, i.e. $d_i=1$. Thus, the tensor network gauge suggests a particular Hamiltonian gauge of the corresponding string net model. The same applies in the fermionic setting. 

The Hamiltonian in Ref.~\cite{GuWen} is given in a specified gauge ($\beta=1$). To obtain the gauge compatible with a tensor network description we apply the transformation 
\begin{equation}
c^\dag \mapsto \frac{1}{\sqrt{\beta}} c^\dag  , \qquad 
c \mapsto \sqrt{\beta} c \;. \label{gauge}
\end{equation}
to ungauge the Hamiltonian and then choose $\beta=\mathrm -i$ which yields the gauge with trivial closed loop factors. The matrix elements in tensor-network gauge of the plaquette operator written as
\begin{align}
Q_p =& \sum_{g_1,\ldots,g_6} p(g_1,\ldots,g_6) \mathcal{F}_p(g_1,\ldots,g_6) \\
& \times \ket{g_1\oplus 1,\ldots, g_6 \oplus1}\bra{g_1,\ldots,g_6} \;, \nonumber
\end{align}
where $\oplus$ denotes addition modulo two are given in Table~\ref{table_hamiltonian}. 

\onecolumngrid

\begin{table*}[t]
 \centering
 \begin{tabular}{ |c|c|c|c|}
\hline
$ i,j,k,l,m,n $ & $p(i,j,k,l,m,n) \mathcal{F}_p(i,j,k,l,m,n)$
& $ i,j,k,l,m,n $ & $p(i,j,k,l,m,n) \mathcal{F}_p(i,j,k,l,m,n)$
\\
\hline
000000 & $ - \alpha/\beta   c_3^\dagger c_6^\dagger$ &
&  \\
\hline
100000 & $c_3^\dagger c_1$ &
010000 & $1/\beta^2 c_1^\dagger c_2^\dagger c_3^\dagger c_6^\dagger    $ \\
001000 & $c_6^\dagger c_2$ &
000100 & $c_6^\dagger c_4$ \\
000010 & $-1/\beta^2 c_3^\dagger c_4^\dagger c_5^\dagger c_6^\dagger $ &
000001 & $ c_3^\dagger c_5$ \\
\hline
110000 & $-1/\beta c_2^\dagger c_3^\dagger$ &
011000 & $-1/\beta c_1^\dagger c_6^\dagger $ \\
001100 & $-\beta c_6^\dagger c_4 c_3 c_2$ &
000110 & $1/\beta c_5^\dagger c_6^\dagger$ \\
000011 & $-1/\beta c_3^\dagger c_4^\dagger $ &
100001 & $\beta c_3^\dagger c_6 c_5  c_1$ \\
\hline
101000 & $-\alpha \beta c_2 c_1$ &
010100 & $ \alpha/\beta c_1^\dagger c_2^\dagger c_6^\dagger c_4   $ \\
001010 & $ -\alpha/\beta c_4^\dagger c_5^\dagger c_6^\dagger c_2$ &
000101 & $ \alpha\beta c_5 c_4$ \\
100010 & $ -\alpha/\beta c_3^\dagger c_4^\dagger c_5^\dagger c_1 $ &
010001 & $ \alpha/\beta c_1^\dagger c_2^\dagger c_3^\dagger c_5$ \\
\hline
100100 & $ -\alpha \beta c_4 c_1$ &
010010 & $ \alpha/\beta^2 c_1^\dagger c_2^\dagger c_3^\dagger c_4^\dagger c_5^\dagger c_6^\dagger $ \\
001001 & $\alpha\beta c_5 c_2$ & &\\
\hline
000111 & $1$ &
001110 & $- c_5^\dagger c_6^\dagger c_3 c_2 $ \\
011100 & $c_1^\dagger c_6^\dagger c_4 c_3 $ & &\\
\hline
101100 & $ \alpha \beta^2 c_4 c_3 c_2 c_1$ &
010110 & $ \alpha / \beta^2 c_1^\dagger c_2^\dagger c_5^\dagger c_6^\dagger $ \\
001011 & $ -\alpha c_4^\dagger c_2$ &
100101 & $ -\alpha \beta^2 c_6 c_5 c_4 c_1$ \\
110010 & $ \alpha/ \beta^2 c_2^\dagger c_3^\dagger c_4^\dagger c_5^\dagger $ &
011001 & $ -\alpha c_1^\dagger c_5$ \\
\hline
010101 & $ - c_4^\dagger c_5^\dagger c_2 c_1$ &
&  \\
\hline
\end{tabular}
 \caption{The first $32$ matrix element of the plaquette operator $Q_p$ defined on a hexagon with physical spins $i,j,k,l,m,n$ after ungauging the Hamiltonian given in Ref.~\cite{FermionicToricCode}. The remaining 32 matrix elements follow by Hermitian conjugation.}
\label{table_hamiltonian}
\end{table*}

\end{document}